\documentclass[a4paper,10pt]{article}
\usepackage[xdvi]{graphics}
\usepackage{amssymb}
\usepackage{latexsym}
\usepackage{amsmath}
\usepackage{epsf}
\def\eqalign#1{\,\vcenter{\openup1\jot\ialign
    {\strut\hfil$\displaystyle{##}$&$\displaystyle{{}##}$
      \hfil\crcr#1\crcr}}\,}

\numberwithin{equation}{section}

\title{
A Relativistic Description of Gentry's \\
New Redshift Interpretation
}
\author{
Takakuni SHIMIZU,
Kazuya WATANABE\\
{\em Department of Physics,~Niigata University, Niigata 
950-2181, Japan}}

\date{}

\begin{document}
\maketitle

%
%
\begin{abstract}
We obtain a new expression of the Friedmann-Robertson-Walker metric,
which is an analogue of a static chart of the de Sitter space-time.
The reduced metric contains two functions, $M(T,R)$ and $\Psi(T,R)$,
which are interpreted as, respectively, the mass function and 
the gravitational potential. 
We find that, near the coordinate origin,
the reduced metric can be approximated 
in a static form and that the approximated metric function, $\Psi(R)$,
satisfies the Poisson equation.  
Moreover, 
when the model parameters of the Friedmann-Robertson-Walker metric
are suitably chosen,
the approximated metric coincides with exact solutions
of the Einstein equation with the perfect fluid matter.
We then solve the radial geodesics on the approximated space-time
to obtain the distance-redshift relation of geodesic sources
observed by the comoving observer at the origin.
We find that the redshift is expressed in terms of
a peculiar velocity of the source and 
the metric function, $\Psi(R)$, evaluated at the source position,
and one may think that this is a new interpretation of
{\it Gentry's new redshift interpretation}. 

\end{abstract}

\

%
%
\section{Introduction}

\

Recently, Gentry proposed his new interpretation of the cosmological
redshift and asserted that its origin might be the gravitational potential
rather than the cosmic expansion\cite{1}. 
Though one immediately
finds several shortcomings of his model\cite{2}, we will show that
Gentry's idea can be partially 
applied to some class of cosmological models,
e.g., de Sitter space-time, in a general relativistic manner
to obtain the distance-redshift relation.
The distance-redshift relation is coordinate-independent
(gauge invariant\cite{3}), however, depends on
both the observer and sources. 
Hereafter, let us assume that the observer and sources
are described by timelike geodesics.
For example, the de Sitter space-time
has at least 4 natural charts (coordinate systems), namely 
an open chart, a flat chart, a closed chart and a static chart.
Since there is a family (a vector field) of the comoving geodesics
in the former 3 charts, we identify the comoving geodesics
with the observer and sources associated with each chart. 
Though the comoving geodesic family does not exist
in the static chart, there is a unique family of the geodesics
in which the comoving geodesic observer is included.
In this way, we have the unique correspondence of a chart to
the observer and sources.
In the former 3 charts, the redshift is naturally interpreted
as a consequence of the cosmic expansion in contrast
to the case in the static chart in which the redshift
appears as the gravitational Doppler effect.
The later gravitational Doppler effect may be
interpreted as {\it Gentry's new redshift interpretation}. 

In this paper, motivated by the above example in the de Sitter space-time,
we find a new expression of the Friedmann-Robertson-Walker (FRW) metric,
which is an analogue of a static chart of the de Sitter space-time.
The reduced metric contains two functions, $M(T,R)$ and $\Psi(T,R)$,
which are interpreted as, respectively, the mass function and 
the gravitational potential. 
After giving the precise reformulation of Gentry's new redshift
interpretation, we try to apply our method to
the FRW metric in the following way.
First, because of a complicated form of the reduced metric,
we approximate the metric near the coordinate origin, R=T=0, and
find that the approximated metric becomes 
the static one and that the approximated metric function, $\Psi(R)$,
satisfies the Poisson equation.  
For the reason of that, we will refer to both $\Psi(T,R)$ and 
its approximated form, $\Psi(R)$, as the gravitational potential. 
We then solve the radial geodesics on the approximated space-time
to obtain the distance-redshift relation.
We find that the redshift is expressed in terms of
a peculiar velocity of the source and the gravitational potential,
$\Psi(R)$, evaluated at the source position,
and one may think that this is a correct, general relativistic
description of {\it Gentry's new redshift interpretation}. 

Throughout this paper, we adopt the unit such that $c=G=1$.

\

\section{The Newtonian chart of the Friedmann-Robertson-\\
Walker metric}

\

We consider the FRW metric 
whose matter contents are perfect fluid and 
vacuum energy. We assume the equation of state,
$P=(\gamma-1)\rho$, for the perfect fluid matter, where
$\gamma$ is a constant. For the vacuum energy,
the relation, $P_V=-\rho_V$, holds.
The FRW metric is then written in the following form,
\begin{equation}
\eqalign{
&ds^2=-dt^2+a(t)^2\left[
\frac{dr^2}{1-Kr^2}+r^2d\Omega^2
\right],~~~d\Omega^2=d\theta^2+\sin^2\theta~d\phi^2,\cr
&\left(\frac{\dot{a}}{a}\right)^2
=H_0^2\left[\Omega_0\left(\frac{a_0}{a}\right)^{3\gamma}
-{k}\left(\frac{a_0}{a}\right)^2
+\Omega_V\right],\cr
&
H_0=\frac{\dot{a}_0}{a_0},~\Omega_0=\frac{8\pi\rho_0}{3H_0^2},~
\Omega_V=\frac{8\pi\rho_V}{3H_0^2},~
k=\frac{K}{a_0^2H_0^2}=\Omega_0+\Omega_V-1,
\cr
}
\label{a1}
\end{equation}
where the suffix, $0$, denotes quantities evaluated at the present
time, $t_0$. The parameters, $H_0$, $\Omega_0$ and $\Omega_V$, 
are usually referred to as, respectively, the Hubble parameter,
the density parameter (of matter) and 
the (normalized) cosmological constant.
Another useful cosmological parameter is the deceleration parameter,
$q_0$, defined by
\begin{equation}
q_0=-\frac{\ddot{a_0}}{a_0H_0^2}
=\left(\frac{3}{2}\gamma-1\right)\Omega_0-\Omega_V.
\label{a2}
\end{equation}

Note that any spherically symmetric metric is locally written
in the following form,
\begin{equation}
ds^2=-\left[1+2\Psi(T,R)\right]dT^2+
\left[1-\frac{2M(T,R)}{R}\right]^{-1}dR^2+R^2 d\Omega^2.
\label{a3}
\end{equation}
We now try to express the FRW metric in the above form,
because, as is the case of the Schwarzschild space-time,
the metric functions, $M(T,R)$ and $\Psi(T,R)$,
are interpreted as, respectively, the mass function and
the gravitational potential. 
We first introduce new coordinates, $T'$ and $R$, 
related to $t$ and $r$ by
\begin{equation}
\eqalign{
&T'=\frac{1}{H_0}\left[F(y)+\frac{1}{(-2k)}\log(1-Kr^2)\right],~~
R=ar,\cr
&y=\frac{a}{a_0},~~\frac{dF}{dy}=\left[
\Omega_0 y^{3(1-\gamma)} - k y + \Omega_V y^3\right]^{-1}.\cr
}
\label{a4}
\end{equation}
The metric ($\ref{a1}$) is then expressed in terms of 
$T'$ and $R$ as
\begin{equation}
ds^2=
\frac{{\displaystyle
-\left[1-k\left(\frac{a_0}{a}\right)^2H_0^2R^2\right]
\left[\Omega_0\left(\frac{a}{a_0}\right)^{4-3\gamma}
-k\left(\frac{a}{a_0}\right)^2
+\Omega_V\left(\frac{a}{a_0}\right)^4\right]{dT'}^2
+dR^2        }}
       {{\displaystyle
1-\left[\Omega_0\left(\frac{a_0}{a}\right)^{3\gamma}
+\Omega_V\right]H_0^2R^2
       }}
+ R^2d\Omega^2.
\label{a5}
\end{equation}

One finds that the comoving geodesic observer, 
i.e., $r=0$ in the Robertson-Walker (RW) chart ($\ref{a1}$),
is mapped to $R=0$ and that  
a proper time, $T$, of the observer is given by 
\begin{equation}
dT^2=\left[\Omega_0\left(\frac{a}{a_0}\right)^{4-3\gamma}
-k\left(\frac{a}{a_0}\right)^2
+\Omega_V\left(\frac{a}{a_0}\right)^4\right]_{R=0}{dT'}^2.
\label{a6}
\end{equation}
When one adopts the new time coordinate, $T$, defined above,
the metric is reduced to the standard form ($\ref{a3}$),
which is referred to as the Newtonian chart, hereafter.
We omit an explicit expression of 
the gravitational potential, $\Psi(T,R)$, because of
its complicated form. 
On the other hand, the mass function, $M(T,R)$, takes
the following simple form,
\begin{equation}
M(T,R)
=\frac{1}{2}\left[\Omega_0\left(\frac{a_0}{a}\right)^{3\gamma}
+\Omega_V\right]H_0^2R^3.
\label{a7}
\end{equation}
It is important to note
that the reduced metric ($\ref{a5}$) has a coordinate singularity, 
$R=2M(T,R)$, and we find that it coincides with the cosmological
apparent horizon. That is, the expansion of the radial ingoing
null geodesic congruences vanishes along 
the trajectory, $R=2M(R,T)$.
Moreover, we have other coordinate singularities,
\begin{equation}\eqalign{
1-k\left(\frac{a_0}{a}\right)^2H_0^2R^2=0~~&\leftrightarrow~~
r=\frac{1}{\sqrt{K}},\cr
\Omega_0\left(\frac{a_0}{a}\right)^{3\gamma}
-k\left(\frac{a_0}{a}\right)^2+\Omega_V=0~~&\leftrightarrow~~
\dot{a}=0,
\cr}
\label{a8}
\end{equation}
which indicate the maximum values
of $r$ and $a(t)$ for the spatially closed ($K>0$)
cosmological models.

\

\section{The chart-associated observer and the redshift}

\

Though the distance-redshift relation is chart-independent,
it crucially depends on both the observer and sources.
As a simple example, the distance-redshift relation
in the Minkowski space-time and that in the Milne universe
are compared in the following.
In the Minkowski space-time, the metric is given by
\begin{equation}
ds^2=-dT^2+dR^2+R^2d\Omega^2,~~\rightarrow~~\Psi(T,R)=M(T,R)=0,
\label{b1}
\end{equation}
and the radial geodesic tangent is given by
\begin{equation}
{\rm Timelike}:~U^a=(\sqrt{1+v^2},v,0,0),~~~~
{\rm Ingoing~ null}:~K^a=(1,-1,0,0),
\label{b2}
\end{equation}
where $v$ is a constant along each timelike 
geodesic and denotes a peculiar velocity of the source
in the Newtonian chart (\ref{b1}).
The redshift, $z$, measured by the comoving observer is given by
\begin{equation}
1+z=v+\sqrt{1+v^2},
\label{b3}
\end{equation}
and $z=0$ for the comoving sources, $v=0$.
In the Milne universe, the metric is given by
\begin{equation}
ds^2=-dt^2+a(t)^2\left(
\frac{dr^2}{1+r^2}+r^2d\Omega^2\right),~~a(t)=t>0,
\label{b4}
\end{equation}
and the radial geodesic tangent is given by
\begin{equation}
\eqalign{
&{\rm Timelike}:~u^a=\left(\sqrt{1+\beta^2\left(\frac{a_0}{a}\right)^2},
\frac{a_0\beta}{a^2}\sqrt{1+r^2},0,0\right),\cr
&{\rm Ingoing~ null}:~k^a=\left(\frac{1}{a}
,-\frac{\sqrt{1+r^2}}{a^2},0,0\right),\cr
}
\label{b5}
\end{equation}
where $\beta$ is a constant along each timelike geodesic
and denotes a peculiar velocity of the source
in the RW chart (\ref{b4}).
The redshift, $z$, measured by the comoving observer is given by
\begin{equation}
1+z=\frac{a_0}{a}\left[\beta\frac{a_0}{a}+
\sqrt{1+\beta^2\left(\frac{a_0}{a}\right)^2}\right].
\label{b6}
\end{equation}
Note that, even when the sources are comoving, i.e., $\beta=0$,
the redshift does not vanish.
The Milne universe consists of all radial timelike geodesics, $R(T)$,
with the following boundary conditions,
\begin{equation}
T>0,~~\lim_{T\rightarrow0}R(T)=0,
~\lim_{T\rightarrow0}\frac{dR}{dT}(T)=\frac{v}{\sqrt{1+v^2}},~
0\leq \forall v<\infty,
\label{bb}
\end{equation}
in the Newtonian chart.
One will find that the peculiar velocity, $v$,
in the Newtonian chart is related to the scale factor,
$a(t)$, as
\begin{equation}
\frac{a_0}{a}=v+\sqrt{1+v^2}.
\label{b7}
\end{equation}
That is, the redshift due to the peculiar velocity
in the Minkowski space-time is interpreted as
the redshift due to the cosmic expansion
in the Milne universe.
This simple example demonstrates that the comoving geodesics
associated with different charts do not necessarily coincide each other.

Another interesting demonstration can be done
in the de Sitter space-time\cite{4}. In the Newtonian chart,
the metric is given by
\begin{equation}
\eqalign{
ds^2&=-(1-H^2R^2)dT^2+(1-H^2R^2)^{-1}dR^2+R^2d\Omega^2,~
H=\sqrt{\frac{8\pi\rho_V}{3}},\cr
&\rightarrow~~
\Psi(T,R)=-\frac{1}{2}H^2R^2,~~M(T,R)=\frac{1}{2}H^2R^3,
\cr}
\label{b8}
\end{equation}
and the radial geodesic tangent is given by
\begin{equation}
{\rm Timelike}:~U^a=\left([1-H^2R^2]^{-1},\pm H R,0,0\right),~~~
{\rm Ingoing~ null}:~K^a=\left([1-H^2R^2]^{-1},-1,0,0\right),
\label{b9}
\end{equation}
where we chose an integration constant such that
$U^a$ describes a family of the timelike geodesics 
in which the comoving geodesic observer, $R=0$, is included.
Up to the sign of $U^1$, the above choice determines
the unique family of geodesics, which are interpreted 
as the observer and sources. 
We take a positive sign of $U^1$ 
and obtain the following angular diameter distance-redshift
relation\cite{30} in the Newtonian chart,
\begin{equation}
R=\frac{1}{H}\frac{z}{1+z}.
\label{b10}
\end{equation}
Hereafter, the angular diameter distance is referred to as
the distance for simplicity.
Accordingly, one has
\begin{equation}
z=\frac{\sqrt{-2\Psi(T,R)}}{1-\sqrt{-2\Psi(T,R)}}.
\end{equation}
When $z<<1$, one has Hubble's law, $z\sim\sqrt{-2\Psi(R)}=HR$, and 
$H$ can be identified with the Hubble parameter
measured by the comoving observer in the Newtonian chart.
When we take a negative sign of $U^1$, 
we have the distance-blueshift relation. 
Similarly, one will find the following distance-redshift relation of the
comoving sources in the RW chart parametrized by $q_0$,
\begin{equation}
R=\frac{1}{H}\left[
1-\frac{1}{1+z}\sqrt{-q_0+(1+q_0)(1+z)^2}
\right]\approx\frac{\sqrt{-q_0}}{H}z=\frac{z}{H_0}~~{\rm for}~z<<1.
\label{b11}
\end{equation}
Note that
the observed Hubble parameter, $H_0$, is chart-dependent
and that, when $q_0=-1$, the distance-redshift relation 
coincides with that in the Newtonian chart. 
Therefore, when $q_0=-1$, the same 
distance-redshift relation is interpreted
in two different manners, namely the redshift
in the RW chart
due to the cosmic expansion, $a(t)$, and 
the redshift in the Newtonian chart
due to the gravitational potential, $\Psi(T,R)$.

\

\section{Revising of Gentry's new redshift interpretation}

\

We now try to interpret the cosmological redshift
in terms of the gravitational potential, $\Psi(T,R)$.
In doing that, we have a problem that
we cannot obtain an explicit form of
the function, $F(y)$, appearing in $(\ref{a4})$.
However, since the Newtonian interpretation
of the metric functions, $M(T,R)$ and $\Psi(T,R)$,
is useful only when the gravitational potential, $\Psi(T,R)$, is small. 
We therefore approximate the metric functions, 
$M(T,R)$ and $\Psi(T,R)$, 
near the coordinate origin, R=T=0, corresponding to
$r=0$, $t=t_0$ in the RW chart,
and will find that the approximated metric is
in a static form and that the approximated gravitational 
potential, $\Psi(R)$, satisfies the Poisson equation.  
We will then interpret the cosmological redshift (Hubble's law)
in terms of the gravitational potential, $\Psi(R)$. 

Assuming that $H_0|t_0-t|$ and $H_0a_0r$ are equally small
quantities, we drop their cubic terms.
One immediately finds that
\begin{equation}
y=\frac{a(t)}{a_0}=1-H_0\Delta t-\frac{1}{2}q_0(H_0\Delta t)^2
\equiv1-\Delta y,
\label{c1}
\end{equation}
where $\Delta t=t_0-t$.
Then we have
\begin{equation}
\eqalign{
H_0T'
&=F(y)+\frac{1}{(-2k)}\log(1-Kr^2)\cr
&=F(1)-\frac{dF}{dy}(1)\Delta y+\frac{1}{2}\frac{d^2F}
{dy^2}(\Delta y)^2+\frac{1}{2}H_0^2R^2\cr
&=F(1)-H_0\Delta t+\frac{q_0-1}{2}(H_0\Delta t)^2
+\frac{1}{2}H_0^2R^2,\cr
}\label{c2}
\end{equation}
which is solved for $\Delta t$ as
\begin{equation}
H_0\Delta t=H_0\Delta T'+\frac{q_0-1}{2}(H_0\Delta T')^2
+\frac{1}{2}H_0^2R^2,
\label{c3}
\end{equation}
where $H_0\Delta T'=F(1)-H_0T'$. Moreover, $y$ 
is expressed in terms of $\Delta T'$ and $R$ as
\begin{equation}
y=1-H_0\Delta T'-\frac{2q_0-1}{2}(H_0\Delta T')^2-
\frac{1}{2}H_0^2R^2.
\label{c4}
\end{equation}
With ($\ref{c4}$),
we now have an approximate of the metric ($\ref{a5}$) as
\begin{equation}
ds^2=
-\left[1+Q(\Delta T')+2\Psi(R)
\right](d\Delta T')^2
+\left[1-\frac{2M(R)}{R}\right]^{-1}dR^2+R^2d\Omega^2,
\label{c5}
\end{equation}
where
\begin{equation}
\eqalign{
&\Psi(R)=\frac{1}{2}q_0H_0^2R^2,~~
M(R)=\frac{1}{2}(\Omega_0+\Omega_V)H_0^2R^3,\cr
&Q(\Delta T')=2(q_0-1)H_0\Delta T'+\cr
&~~~\left[
2+\frac{(3\gamma-8)(3\gamma-2)}{2}\Omega_0+8\Omega_V
+\frac{(3\gamma-2)^2}{2}\Omega_0^2
+(2-3\gamma)\Omega_0\Omega_V+2\Omega_V^2
\right](H_0\Delta T')^2.
\cr
}\label{c6}
\end{equation}
A proper time, T, of the comoving observer, $R=0$,
is given by
\begin{equation}
T=-\int_{0}^{\Delta T'}\sqrt{1+Q(\tau)}d\tau,
\label{c7}
\end{equation}
and we finally obtain the approximated metric as
\begin{equation}
ds^2=
-\left[1+2\Psi(R)\right]dT^2
+\left[1-\frac{2M(R)}{R}\right]^{-1}dR^2+R^2d\Omega^2.
\label{c8}
\end{equation}

We find that the gravitational potential, $\Psi(R)$,
satisfies the following Poisson equation,
\begin{equation}
\nabla^2\Psi=3q_0H_0^2=4\pi(\rho_0+3P_0+\rho_V+3P_V),
\label{c9}
\end{equation}
which can be identified with the Newtonian approximation of
the $(0,0)$-component of the Einstein equation,
\begin{equation}
R_{00}=8\pi\left(T_{00}-\frac{T^a_a}{2}g_{00}\right).
\label{c10}
\end{equation}
The metric function, $M(R)$, can be identified with
the mass function as follows,
\begin{equation}
\frac{dM}{dR}=\frac{3}{2}(\Omega_0+\Omega_V)H_0^2R^2
=4\pi R^2(\rho_0+\rho_V).
\label{c11}
\end{equation}
Moreover, we find a further physical role of $\Psi(R)$ and $M(R)$
as follows. A tangent vector, $T^a$, of the apparent horizon trajectory
(see Section 2), $x_{\rm AH}^{a}(t)$, 
is given by $T^a=\partial x_{\rm AH}^{a}/\partial t$ in 
the RW chart. A norm of $T^a$ at $t_0$ is given by
\begin{equation}
\eqalign{
&r_{\rm AH}(t_0)=\frac{1}{H_0a_0}(\Omega_0+\Omega_V)^{-\frac{1}{2}},~~
\left[T^a\right]_{t_0}=\left(1,
\frac{q_0}{a_0}(\Omega_0+\Omega_V)^{-\frac{3}{2}},0,0\right),\cr
\rightarrow~~
&\left[g_{ab}T^aT^b\right]_{t_0}=-1+\frac{q_0^2}{(\Omega_0+\Omega_V)^2}
=-1+\left[\frac{\Psi(R)}{M(R)}R\right]^2,\cr
}
\label{cc1}
\end{equation}
without any approximation.  
That is, if the present time is a mass-dominant era,
$|M(R)|>|R\Psi(R)|$,
the apparent horizon trajectory is timelike.

It may be interesting to note that, though the metric
(\ref{c8}) is derived as an approximated form of
the FRW metric, it coincides with exact solutions
of the Einstein equation with perfect fluid matter
in the following particular cases.

\begin{itemize}
\item{$\gamma \Omega_0=0$} \\
In this case, we have the de Sitter space-time for 
$q_0<0$
and the anti-de Sitter space-time for $q_0>0$.
\item{$q_0=0$} \\
In this case, we have Einstein's static universe
with $\rho_V=(3\gamma/2-1)\rho_0$. 
The same metric is also interpreted as a solution
of the Tolman-Oppenheimer-Volkoff equation\cite{5,6}
with the equation of state,
\begin{equation}
P_{\rm eff}=-\frac{1}{3}\rho_{\rm eff},~~
\rho_{\rm eff}=\rho_0+\rho_V={\rm constant}.
\label{c12}
\end{equation}
Note that the perfect fluid with 
$(\rho_{\rm eff},P_{\rm eff})$ satisfies
all the energy conditions, i.e., the weak energy condition,
the strong energy condition and the dominant energy condition. 
\end{itemize}

Let us obtain the distance-redshift relation
on the approximated space-time ($\ref{c8}$).
We will show below that the redshift
is independent of the mass function, $M(R)$. 
The radial geodesic tangent is given by
\begin{equation}
\eqalign{
&{\rm Timelike}:~U^a=\left(\frac{\sqrt{1+v^2}}{1+q_0H_0^2R^2},
\pm\sqrt{\frac{
\left[1-(\Omega_0+\Omega_V)H_0^2R^2\right]
\left[v^2-q_0H_0^2R^2\right]}{1+q_0H_0^2R^2}},0,0\right),\cr
&{\rm Ingoing~ null}:~K^a=
\left(\frac{1}{1+q_0H_0^2R^2},
-\sqrt{\frac{1-(\Omega_0+\Omega_V)H_0^2R^2}{1+q_0H_0^2R^2}},0,0\right),\cr
}\label{c13}
\end{equation}
where $v$ is a constant along the geodesic 
and denotes a peculiar velocity of the source.

When $q_0\leq0$ and $v=0$, the distance-redshift relation
of the sources becomes
\begin{equation}
\eqalign{
z&=\frac{\sqrt{-q_0}H_0R}{1-\sqrt{-q_0}H_0R}\cr
&=\frac{\sqrt{-2\Psi(R)}}{1-\sqrt{-2\Psi(R)}}
\approx\sqrt{-2\Psi(R)}~~~{\rm for}~|\Psi(R)|<<1,\cr
\rightarrow~~R&=\frac{1}{\sqrt{-q_0}H_0}\frac{z}{1+z}
\approx\frac{z}{\sqrt{-q_0}H_0}~~~{\rm for}~z<<1,\cr}
\label{c14}
\end{equation}
and the comoving observer in the approximated Newtonian chart ($\ref{c8}$)
measures an effective Hubble parameter, $H_{\rm eff}=\sqrt{-q_0}H_0.$
In the de Sitter space-time,
the above results hold exactly.

On the other hand, when $q_0>0$, no family of 
the comoving geodesics ($v=0$) exists. 
In this case, the redshift, $z$, measured by the comoving observer,
$R=0$, is given by
\begin{equation}
1+z=\frac{\sqrt{1+v^2}+\sqrt{v^2-q_0H_0^2R^2}}{1+q_0H_0^2R^2}
=\frac{\sqrt{1+v^2}+\sqrt{v^2-2\Psi(R)}}{1+2\Psi(R)},
\label{c15}
\end{equation}
and the distance-redshift relation {\it formally} becomes 
\begin{equation}
\eqalign{
R&=\frac{1}{H_0\sqrt{q_0}(1+z)}\sqrt{-1+2(1+v^2)^{\frac{1}{2}}
(1+z)-(1+z)^2}\cr
&\approx \frac{1}{H_0\sqrt{q_0}}\left(1-\frac{z}{2}\right)
\sqrt{2(1+v^2)^{\frac{1}{2}}-2}
~~{\rm for}~z<<1.\cr}
\label{c16}
\end{equation}
It should be noted that $z$ becomes negative 
when $v$ is smaller than the R-dependent critical value,
$v_{\rm c}=\sqrt{2\Psi(R)+\Psi(R)^2}\approx\sqrt{2\Psi(R)}$,
and that, in contrast to the previous negative $q_0$ case,
we {\it formally} have the following tilted Hubble law,
\begin{equation}
R=R_0(v)+\hat{H}_0(v)z,~~
R_0(v)=\frac{
\sqrt{2(1+v^2)^{\frac{1}{2}}-2}}
{H_0\sqrt{q_0}},~~
\hat{H}_0(v)=-\frac{
\sqrt{(1+v^2)^{\frac{1}{2}}-1}}
{H_0\sqrt{2q_0}}.
\label{c17}
\end{equation}
One may think that this tilted Hubble law is inconsistent 
with Hubble's law, $z=H_0R$.
However, it is not necessarily the case because
the integration constant, $v$, is constant along
each geodesic and is not necessarily a global constant.
For example, the comoving geodesic, $u^a=\delta^a_0$,
in the closed RW chart of the de Sitter space-time
has its components, $U^a$, in the Newtonian chart parametrized by
\begin{equation}
v(T,R)=\frac{HR}{\sqrt{\cosh^2HT-H^2R^2\sinh^2HT}},~~~
U^a\partial_a v=0.
\label{c18}
\end{equation}
That is, the ``{\it constant}'' depends on both $T$ and $R$.
Similarly, the comoving geodesic, $u^a=\delta^a_0$,
in the open RW chart of the anti-de Sitter space-time,
\begin{equation}
ds^2=-dt^2+\sin^2Ht\left(\frac{dr^2}{1+H^2r^2}+r^2d\Omega^2\right),~~~
H^2=-\frac{8}{3}\pi\rho_V>0,
\label{c19}
\end{equation}
has its components, $U^a$, in the Newtonian chart,
\begin{equation}
ds^2=-(1+H^2R^2)dT^2+(1+H^2R^2)^{-1}dR^2+R^2d\Omega^2,
\label{c20}
\end{equation}
parametrized by the ``{\it constant}'',
\begin{equation}
v(T,R)=\frac{HR}{\sqrt{\sin^2HT-H^2R^2\cos^2HT}},~~~
U^a\partial_a v=0.
\label{c21}
\end{equation}
For generic geodesics in the Newtonian chart, one can adopt
different values of $v$, however, the choice
of globally vanishing $v$ is forbidden when $q_0>0$. 

Finally, we briefly discuss the Newtonian interpretation of
the integration constant, $v$.
In the slow motion limit, the radial
timelike geodesic is approximated as
\begin{equation}
\eqalign{
&\frac{d^2R}{dT^2}
=-\frac{\partial\Psi}{\partial R}=-q_0H_0^2R,\cr
\rightarrow~~~&\left(\frac{dR}{dT}\right)^2
+q_0H_0^2R^2=v^2,~~~~v={\rm integration~constant},\cr
\rightarrow~~~&\frac{dR}{dT}=\pm\sqrt{v^2-q_0^2H_0^2R^2},\cr
}
\label{c22}
\end{equation}
which is compared with ($\ref{c13}$).
One finds that when $q_0>0$, $v^2$ plays a role of 
a conserved total energy of the harmonic oscillator
and, therefore, does not vanish unless $R=dR/dT=0$.

\

\section{Summary and discussion}

\

In this paper, we have obtained a new expression of
the FRW metric in the Newtonian chart and
have found that the apparent horizon appears
as a coordinate singularity. The appearance of
this coordinate singularity may be understood in the following way.
Let $u$ and $v$ be null coordinates such that
ingoing null geodesics are represented by the equation,
$u=$constant, and that $v$ is an affine parameter
of the ingoing null geodesic. Since the geodesically
incompleteness indicates the initial singularity
in the FRW metric, $v$ does not have any coordinate singularity.
When the further coordinate transformation,
$v\rightarrow R$, is done, we have
a coordinate singularity if the coordinate
transformation becomes singular, i.e.,
$|\partial R/\partial v|\rightarrow0$ or $\infty$, and
the previous condition, $\partial R/\partial v=0$,
is equivalent with the existence of
the apparent horizon.  
After interpreting the redshift
in the Milne universe and the de Sitter space-time in Gentry's manner,
we have tried to obtain the distance-redshift relation
in the approximated Newtonian chart
in terms of the Newtonian potential, $\Psi(R)=q_0H_0^2R^2/2$,
and the peculiar velocity, $v$, of the source.
We have found that the sign of
the deceleration parameter, $q_0$,
significantly affects the redshift interpretation in 
the Newtonian chart. 
That is, when $q_0$ is positive, we have no natural family 
of the timelike geodesics in the approximated
Newtonian chart in contrast to the negative $q_0$ cases.

Now we discuss the implication of our results in the numerical
relativity. Note that one often adopts particular coordinates
in the numerical relativity, e.g., the maximal slice\cite{7},
in order to avoid the numerical divergence due to
the singularities. In such numerically adopted charts,
even the well-known metric will take an unfamiliar form,
and, moreover, one may have unexpected numerical
coordinate singularities. 
It is therefore useful to study how the well-known space-time
is expressed in the unfamiliar charts\cite{7}.
As far as an empty or approximately empty space-time is concerned,
one may have comparatively few difficulties
in interpreting the numerical results.
For example, the appearance of the null apparent horizon
will be interpreted as a strong evidence
of the occurrence of the de Sitter expansion,
e.g., the inflationary era\cite{8},
in numerically studying the dynamical and global evolution
of the early universe. 
However, when the existence of the matter cannot be
negligible, it is difficult
to distinguish the isotropic, homogeneous space-time
from generic isotropic, inhomogeneous space-time
in the Newtonian chart, as has been shown in this paper.
 
In order to examine further properties
of the apparent horizon,
we calculate a trace of the extrinsic curvature
and find that
\begin{equation}
K^a_a=-3H_0
\sqrt{
\Omega_0\left(\frac{a_0}{a}\right)^{3\gamma}-
k\left(\frac{a_0}{a}\right)^2+\Omega_V},
\label{d1}
\end{equation}
in the FW chart and that
\begin{equation}
{\displaystyle
K^a_a=
\frac{3}{2}\gamma\Omega_0H_0\left(\frac{a_0}{a}\right)^{3\gamma}
H_0^2R^2
\sqrt{
  \frac{
       \left[   \Omega_0\left(\frac{a_0}{a}\right)^{3\gamma}
               -k\left(\frac{a_0}{a}\right)^2+\Omega_V   
       \right]
       \left[
                1-k\left(\frac{a_0}{a}\right)^2H_0^2R^2
       \right]
        }
        {
        \left\{
        1-\left[  \Omega_0\left(\frac{a_0}{a}\right)^{3\gamma}
         +\Omega_V  \right]H_0^2R^2
        \right\}^{3} 
        }
    },
  }
\label{d2}
\end{equation}
in the Newtonian chart. One finds that $K^a_a$ is
{\it homogeneous} in the RW chart
and speculates that the space-time itself is homogeneous 
in contrast to the case in the Newtonian chart
in which $K^a_a$ is {\it inhomogeneous}.
Eardley and Smarr\cite{9} have defined a crushing singularity
as
\begin{equation}
|K^a_a|~~\rightarrow~~\infty~~{\rm uniformly}.
\label{d3}
\end{equation}
In the RW chart, one immediately finds that
$a=0$ is a crushing singularity generically.
On the other side, 
the generic singularity, $a=0$, is
outside the coordinate singularity, $R=2M(T,R)$,
in the Newtonian chart, and
the divergence of $K^a_a$ appears only at
the apparent horizon, $R=2M(T,R)$, unless $\gamma\Omega_0=0$.
This divergence does not occur on the hypersurface, $T=$constant,
in general and is not a crashing singularity.
However, one should remember that the motivation of defining 
the crushing singularity is the conjecture
that the singular behavior of the adopted time function such that
\begin{equation}
|K^i_i|\sim\frac{1}{2}\left|g^{ij}\frac{1}{\sqrt{-g_{00}}}
\partial_0g_{ij}
\right|\rightarrow\infty,
\label{d4}
\end{equation}
may correspond to the appearance of
a singularity. In this sense, the divergence
of $K^a_a$ at the apparent horizon, $R=2M(T,R)$,
may be interpreted as the appearance
of the physical singularity by the observer in the Newtonian chart
though it is not a true singularity.
Moreover, we calculate
a 3-dim scalar curvature, ${\cal R}$, induced on the spacelike hypersurface,
$T=$constant, in the Newtonian chart and find that
\begin{equation}
{\cal R}=6H_0^2\left\{
\Omega_0\left(\frac{a_0}{a}\right)^{3\gamma}+\Omega_V
+\frac{
\gamma\Omega_0\left(\frac{a_0}{a}\right)^{3\gamma}
\left[\Omega_0\left(\frac{a_0}{a}\right)^{3\gamma}
-k\left(\frac{a_0}{a}\right)^2+\Omega_V
\right]H_0^2R^2
}{
1-\left[\Omega_0\left(
\frac{a_0}{a}\right)^{3\gamma}+\Omega_V\right]H_0^2R^2
}
\right\},
\label{d5}
\end{equation}
which diverges at the apparent horizon, $R=2M(T,R)$, unless $\gamma\Omega_0=0$.
That is, the apparent horizon 
in the Newtonian chart almost always appears
not only as a coordinate singularity but also
as {\it an apparent singularity} 
such that the extrinsic curvature and the 3-dim curvature
simultaneously diverge unless $\gamma\Omega_0=0$. 

\



\begin{thebibliography}{99}
\bibitem{1}R.V.Gentry,
         Mod.Phys.Lett. $\mathbf{A12}$, 2919, (1997).
\bibitem{2}S.Carlip, R.Scranton,
         astro-ph/9808021, (1998). 
\bibitem{3}M.Sasaki, 
         M.N.R.A.S. $\mathbf{228}$, 653, (1987).
\bibitem{4}S.W.Hawking, G.F.R.Ellis,
        {\it The large scale structure of space-time},
        (Cambridge University Press, 1973).
\bibitem{30}S.Weinberg,
        {\it Gravitation and cosmology},
        (Wiley, 1972).
\bibitem{5}R.C.Tolman,
         Phys.Rev. $\mathbf{55}$, 364, (1939).
\bibitem{6}J.R.Oppenheimer, G.M.Volkoff,
         Phys.Rev. $\mathbf{55}$, 374, (1939).
\bibitem{7}F.Estabrook, et al.,
         Phys.Rev $\mathbf{7}$, 2814, (1973).
\bibitem{8}A.Guth,
         Phys.Rev $\mathbf{D23}$, 347, (1981).
\bibitem{9}D.M.Eardley, L.Smarr,
         Phys.Rev $\mathbf{D19}$, 2239, (1979).
\end{thebibliography}
\end{document}